\documentclass[aip,graphicx]{revtex4-1}    

\usepackage{amsmath}
\usepackage{graphicx}
\usepackage{epstopdf}
\draft 

\begin{document}

\title{Geometric Scaling for a Detonation Wave Governed by a Pressure-Dependent Reaction Rate and Yielding Confinement} 

\author{J. Li}%
\email[]{lijianling@mail.nwpu.edu.cn}
\affiliation{Institute of Fluid Physics\\China Academy of Engineering Physics, Mianyang, China}

\author{X. Mi}
\email[]{xiaocheng.mi@mail.mcgill.ca}

\author{A. J. Higgins}%
\email[Corresponding author: ]{andrew.higgins@mcgill.ca}
\affiliation{Department of Mechanical Engineering\\McGill University, Montreal, Quebec, H3A 2K6 Canada}

\date{\today}

\begin{abstract}
The propagation of detonation waves in reactive media bounded by an inert, compressible layer is examined via computational simulations in two different geometries, axisymmetric cylinders and two dimensional, planar slabs. For simplicity, an ideal gas equation of state is used with a pressure-dependent reaction rate that results in a stable detonation wave structure.  The detonation is initiated as an ideal Chapman-Jouguet (CJ) detonation with a one-dimensional structure, and then allowed to propagate into a finite diameter or thickness layer of explosive surrounded by an inert layer.  The yielding confinement of the inert layer results in the detonation wave decaying to a sub-CJ steady state velocity or failing entirely.  Simulations are performed with different values of the reaction rate pressure exponent ($n = 2$ and $3$) and different impedance confinement (greater than, less than, and equal to that of the explosive).  The velocity decrement and critical dimension (critical diameter or thickness) are determined, and a $2:1$ scaling between the cylinder diameter and slab thickness results is confirmed, in good agreement with curvature-based models of detonation propagation. The measured shock front curvature and detonation velocity relation ($D_\mathrm{N}$-$\kappa$) agrees with the classic model of Wood and Kirkwood. The computational simulations are compared to a simple, analytic model that treats the interaction of the confinement with the detonation products via Newtonian theory and a model that assumes a continuous variation in shock front curvature with the shock angle at the interface with the confinement matching the angle determined by shock polar analysis. The Newtonian model works very well for the case of high impedance confinement, while the shock front curvature model agrees with the simulations for the case of low impedance confinement.
\end{abstract}

\maketitle

\section{Introduction}
\label{sec:introduction}
The study of how detonation waves respond to losses is the primary experimental and theoretical means of understanding detonation dynamics. Specifically, for condensed phase explosives, quantifying the relationship between the diameter of a cylindrical charge with yielding confinement, propagation velocity, and front curvature is the principal technique used in developing models for detonation propagation in a given explosive. Similarly, for gaseous explosives, the response of a detonation propagating in a finite diameter channel or tube with mass, heat, and momentum losses to the walls reveals the link between the dynamic parameters (e.g., detonation cell size) that characterize its structure and its global propagation behavior.\\

In recent years, the scaling between detonations propagating with losses in different geometries has been investigated, with some unanticipated results being generated.  Studies have examined the scaling between propagation in the two most frequently encountered geometries, namely, propagation in a cylindrical tube and in a two-dimensional slab, for both mixtures of detonable gases\cite{Radulescu200229, Radulescu2003, Meredith2010} and condensed phase explosives\cite{Petel2006DS, Petel2007JL, Higgins2009APS, Higgins2013APS, Jackson2013APS}. In classic, front-curvature governed detonations, such as exhibited in so-called “ideal” explosives with yielding confinement, the propagation velocity deficit and critical velocity at failure are expected to scale between the axisymmetric (diameter $d$) and two-dimensional (slab thickness $t$) geometries according to $d:t=2:1$, as shown by a two-dimensional analysis of the detonation structure performed by Bdzil.\cite{Bdzil1981}  In the case of strong confinement, Bdzil showed that the scaling should be exactly $2:1$, while in the case of a weakly confined charge (resulting in greater front curvature), the scaling is somewhat less than $2:1$, meaning that the same detonation velocity would be observed in an explosive stick with a diameter slightly less than twice the thickness of a two-dimensional slab of the same explosive.\cite{Bdzil1981}\\

The experimental results obtained over the last decade have suggested that the picture is considerably more complex than that described by detonation front curvature theory. Radulescu\cite{Radulescu200229, Radulescu2003} used the scaling between gaseous detonations propagating in channels with porous walls in order to investigate the mechanism of failure. In highly argon diluted mixtures which exhibit a weak but regular cellular structure, comparing propagation in two-dimensional channels and circular tubes (both with porous walls) verified an approximately $2:1$ scaling.  The results in the channels were also insensitive to the aspect ratio, suggesting that global front curvature governs the propagation of detonation waves in this class of mixture. In undiluted mixtures with a highly irregular cellular structure, however, this scaling broke down and became sensitive to the aspect ratio of the channel used, suggesting a mode of propagation that relies upon local interactions of the transverse waves that define the cellular structure.\cite{Radulescu200229, Radulescu2003} Similar anomalous scaling has been observed in critical diameter experiments, in which a detonation in a tube or channel emerges into an unconfined space.  For irregular mixtures, this experiment usually generates the well-known result that the critical diameter is $13\mathrm{\lambda}$, where $\lambda$ is the detonation cell size.  Although this transmission problem is transient (as opposed to a steady propagation limit), it has been suggested that curvature-based scaling may still predict the ratio between the two geometries.\cite{Lee2008, Benedick1984} In essence, since a spherical front has a curvature given by $\mathrm{\kappa} = 2/R_\mathrm{c}$, while a cylindrical front has curvature $\mathrm{\kappa} = 1/R_\mathrm{c}$ (where $R_\mathrm{c}$ is the radius of curvature), the expectation in detonation transmission studies was that the critical thickness should be approximately half of the critical diameter, or about $6$-$7\mathrm{\lambda}$. Experiments have shown, however, that the scaling is closer to $4:1$ for most mixtures, resulting in a critical thickness for a detonation emerging from a slot of only $\approx3\mathrm{\lambda}$, a result that to date has defied explanation.\cite{Lee2008} If a highly argon diluted mixture with a weak, regular cellular structure is used instead, an approximately $2:1$ scaling between the critical diameter and slot is recovered, in agreement with curvature considerations.\cite{Meredith2010}\\

The scaling between geometries for the case of condensed-phase (solid or liquid) explosives has been less studied, but the data that exists suggests that, in explosives with fine-scale heterogeneities, the scaling is approximately $2:1$.\cite{Ramsay1985, Gois1996} A study of three different, highly heterogeneous explosives by Petel et al.\cite{Petel2007JL}, however, found that the scaling varied between $2:1$ to $4:1$ as the scale of the heterogeneity increased in comparison to the dimension of the charge. A study by Gois et al.\cite{Gois1996} with a gelled-nitromethane explosive sensitized with glass microballoons (GMB) found a scaling of $2:1$ over a wide range of GMB concentrations, until the concentration of GMB became dilute and a scaling of $d:t=3:1$ was observed, a result that was later experimentally reproduced by Higgins.\cite{Higgins2009APS} Very recently, Jackson and Short have undertaken a thorough measurement of  velocity-diameter and velocity-thickness relations for two plastic-bonded explosives (PBX 9501 and PBX 9502) and a non-ideal blasting compound (ammonium nitrate with fuel oil).\cite{JacksonShort2014} The measured scaling for all three explosives was found to be slightly less than $2:1$ and in excellent agreement with the predictions of the front curvature-based theory of Detonation Shock Dynamics.\cite{Jackson2013APS} The cause of such a wide variation of observed behaviors is uncertain.  Petel et al.\cite{Petel2006DS} suggested that large-scale heterogeneities may result in a break-down in global front curvature and the propagation mechanism becomes dependent upon the local structure of the media. The existence of instabilities (e.g., cellular structure) in condensed-phase explosives might also explain a departure from front curvature-governed behavior, as has been reviewed by Lee\cite{Lee2002} and Higgins\cite{Higgins2013APS}.\\

The purpose of the present study is to computationally investigate the scaling between the two most fundamental geometries for detonation propagation in finite-sized charges via direct numerical simulation.  For simplicity, a single component gas formulation is used for both the explosive medium and the inert confinement.  In order to avoid the complexities of unstable wave structures (i.e., cellular instability) encountered with Arrhenius-based kinetics, a pressure-dependent reaction rate is used.  The resulting detonation behavior is expected to be ideal, controlled by the global curvature of the detonation front.  The main parameter studied is the velocity scaling between the two geometries.  The use of a sufficiently large pressure exponent ($n>2$) results in a critical, turning-point behavior in the steady detonation solution as the charge diameter or thickness is decreased, so that scaling of the critical dimension can also be examined.\\

The results will also be compared to analytic models of propagation with losses due to yielding confinement.  Although a quantitative link between the detonation velocity, shock front curvature, and flow divergence can be made by solving for the reaction zone structure along the centerline, as has been known since the work of Wood and Kirkwood\cite{WoodKirkwood1954, Fickett2000, HigginsChapter2}, this model by itself does not permit the dependence of detonation velocity upon the charge diameter to be directly determined.  It is necessary to provide an additional link between the front curvature and the dimension of the charge and its interaction with the confinement, which is often handled via an \textit{ad hoc} or semi-empirical treatment.\cite{Chan1981} The inclusion of the shock front/wall interaction into the more sophisticated theory of Detonation Shock Dynamics has only recently been undertaken.\cite{Aslam2002, Sharpe2006, BdzilChapter7}\\

For the purposes of this paper, the results of the direct numerical simulations will be compared to and interpreted with the assistance of classical models of detonation propagation.  The simplest of these is a one-dimensional nozzle-flow model where the interaction of the expanding detonation products with the confinement is treated via the Newtonian hypersonic flow model.  The next model derives from a simple geometric construction proposed by Eyring et al.\cite{Eyring1949}, wherein the front curvature is assumed to vary continuously from the axial centerline to match the shock interaction with the confinement, which may be solved via shock polar analysis. The relation between this geometric model and the more sophisticated model of Detonation Shock Dynamics is also established. The details of these models are summarized in the Appendices.  The hierarchy of various models is examined in order to determine the minimum level of sophistication required for a model to correctly capture the dynamics of stable detonation waves and interpret the geometric scaling results.
\section{Problem Statement}
\label{sec:problem}
\begin{figure}
	\centering
		\includegraphics[width=1.0\textwidth]{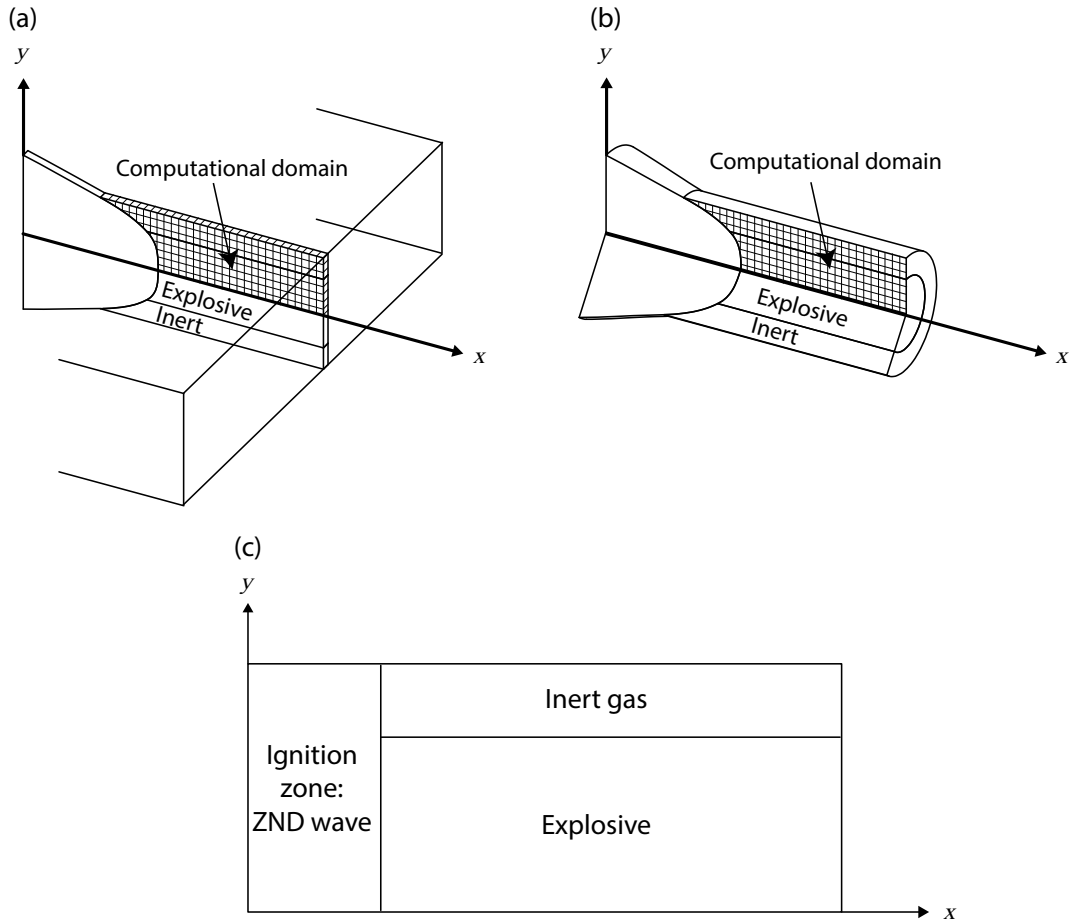}
		\caption{Schematic representation of the problem to be studied showing the computational grid for (a) two-dimensional slab and (b) axisymmetric geometries, and (c) the initial conditions.}
	\label{fig:Figure1}
\end{figure}

The problem examined is detonation propagation in a charge of explosive bounded by inert, yielding confinement, as shown schematically in Fig.~\ref{fig:Figure1}. Two geometries are considered:  planar, two-dimensional charges and cylindrical, axisymmetric charges.  The detonation wave is initialized across the entire transverse span of the domain using a one-dimensional ZND-structured wave propagating at the ideal, Chapman-Jouguet detonation velocity, referred to as the initiation region.  The wave then propagates into the finite thickness or diameter charge bounded by the inert medium.  The detonation undergoes a transient relaxation to the final, steady state propagation velocity, and the terminal propagation velocity is determined.  In the case of a subcritical diameter or thickness, the detonation wave fails entirely, with the reaction zone progressively decoupling from a decaying shock wave. For simplicity, the ideal gas law is assumed:
\begin{equation}
p=\rho\mathrm{R}T
\label{1}
\end{equation}
The reaction rate law is a pressure-dependent reaction rate:
\begin{equation}
\frac{\partial Z}{\partial t}=k(1-Z)\left(\frac{p}{p_\mathrm{CJ}}\right)^n
\label{reactionrate}
\end{equation}
where $Z$ is the reaction progress variable ($Z=0$ for unreacted, $Z=1$ for fully reacted).  A pressure dependent reaction rate, as opposed to the more familiar Arrhenius reaction rate, is used so that the resulting detonation wave is stable.  Two-dimensional detonations governed by Arrhenius reaction rates have been shown to be unstable to transverse perturbations for all values of activation energy.\cite{Short1998, NgChapter3} While we are not aware of a formal stability analysis that has been performed on a pressure-dependent reaction rate of this type, the existence of a stable wave structure will be demonstrated computationally in Section \ref{sec:results}.  Pressure-dependent reaction rates are widely used in modeling solid, polycrystalline explosives and are intended to represent the burn-out of explosive grains from localized ignition centers (so-called hot-spots).\cite{LeeTraver1980} The ignition hot spots and subsequent burn-out of the explosive is a sub-grid phenomenon in most computational simulations of explosives, so the $p^n$ represents mesoscale features that cannot be directly simulated.  The values of $n$ used in models of high explosives are usually in the range $1 \leq n \leq4$.\cite{Cowperthwaite1994, Short2002, Short2008} For pressure exponents of $n>2$, the detonation velocity \textit{vs}. diameter/thickness relation exhibits a turning point behavior that can be associated with the critical dimension at which the detonation fails. For this study, values of $n=2$ (no turning point) and $n=3$ (turning point exists) will be used.\\

The properties of the explosive are chosen to be approximately those of stoichiometric hydrogen/oxygen as follows, $\bigtriangleup q/\mathrm{R}T_1=24$ and $\gamma=1.333$.  A single gas formulation is used for the entire domain (explosives and inert confinement), so that the inert gas layer has identical thermodynamic properties but with the reaction progress variable set to $Z=1$, such that no energy can be released in this layer.  The explosive and inert gases must have the same initial pressure in order to avoid a shock-tube problem in the initial conditions, and this requires that the product of temperature and density for the explosive and confinement must be equal.  The acoustic impedance of the confinement is varied by inversely varying the density and temperature.  Since acoustic impedance is given by $\rho c$ where $c$ is the sound speed, which for an ideal gas $c\sim\sqrt{T}$, maintaining the static pressure as constant while varying the density of a gas results in the impedance varying as $\sqrt{\rho}$. In this study, the density of the confinement is varied from $0.5$ to $10$ $\rho_\mathrm{o}$ in order to vary the impedance of the confinement layer.\\

The use of the ideal gas law coupled with a pressure-dependent reaction rate (usually used for solid explosives) makes this problem a hybrid of gas phase and condensed phase detonation phenomena.    We emphasize that this study does not attempt to model any real explosive system. Rather, the purpose is to examine the simplest possible system that can be easily modelled, and this objective motivated the selection of an ideal gas with a pressure-dependent reaction rate.  Extension of this study to equations of state more representative of actual condensed phase materials (e.g., the stiffened gas equation of state\cite{Short2006}) or reaction rate laws consistent with gas-phase kinetics (Arrhenius kinetics) would be straightforward, but would unnecessarily complicate the analysis for the present purposes.
\section{Numerical Method}
\label{sec:numerical}
The simulations were performed in a lab-fixed reference frame, with the computational grid being extended as the wave propagated to near the end of the domain, and the post-detonation region of the domain being periodically discarded due to computational memory limitations.  The domain was truncated at a time sufficiently long after the passage of the wave so as to occur past the limiting characteristic (with corresponds to the sonic surface for a steady wave) and was verified not to influence the propagation of the wave.\\

The governing unsteady, two-dimensional inviscid Euler equations with the pressure-dependent reaction rate source term described above in Cartesian coordinates and cylindrical coordinates were solved on a uniform computational grid for two-dimensional and axisymmetric geometries respectively. The effects of viscosity, heat conductivity, and diffusion are not taken into account in the simulations reported here. The unsteady governing equations are given as follows:
\begin{equation}
\frac{\partial \vec{U}}{\partial t}+\frac{\partial \vec{E}}{\partial x}+\frac{\partial \vec{F}}{\partial y}+\epsilon{\vec{S}}_{\mathrm{nr}}=\vec{S}_{\mathrm{r}}
\label{N1}
\end{equation}
where the conserved variable vector $\vec{U}$, the inviscid convective flux vectors $\vec{E}$ and $\vec{F}$ in the $x$ and $y$ directions respectively, the axisymmetric cross source terms vector $\vec{S}_{\mathrm{nr}}$, and the chemical reaction source terms vector $\vec{S}_{\mathrm{r}}$ are given by
\begin{equation}
 \vec{U}=
 \begin{pmatrix}
 \rho \\
\rho u \\
\rho v \\
\rho e \\
\rho Z
 \end{pmatrix}
~\vec{E}=
 \begin{pmatrix}
 \rho u \\
\rho u^2+p \\
\rho vu \\
(\rho e+p)u \\
\rho Zu
 \end{pmatrix}
~\vec{F}=
 \begin{pmatrix}
 \rho v \\
\rho uv \\
\rho v^2+p \\
(\rho e+p)v \\
\rho Zv
 \end{pmatrix}
~{\vec{S}}_{\mathrm{nr}}=\frac{1}{r}
 \begin{pmatrix}
 \rho v \\
\rho uv \\
\rho v^2 \\
(\rho e+p)v \\
\rho Zv
 \end{pmatrix}
~{\vec{S}}_{\mathrm{r}}=
 \begin{pmatrix}
 0 \\
0 \\
0 \\
0 \\
\rho {\dot{\omega}}
 \end{pmatrix}
\label{N2}
\end{equation}
where $t$, $\rho$, $p$, $T$, $u$, $v$, $e$, $Z$, and $\dot{\omega}$ represent the time, fluid density, pressure, temperature, velocity in $x$ and $y$ directions, specific total energy, reaction progress variable and reaction rate, respectively. The geometric factor $\epsilon$ equals 0 for two-dimensional flow, and 1 for axisymmetric flow. The specific total energy is given by
\begin{equation}
e=\frac{1}{\gamma-1}\frac{p}{\rho}+\frac{1}{2}(u^2+v^2)-Zq
\label{N3}
\end{equation}
where $q$ is the heat release of chemical reaction per unit mass, and $\gamma$ is the specific heat ratio. The one-step reaction rate $\dot{\omega}$ is given by the pressure-dependent reaction rate model, i.e., Eq.~(\ref{reactionrate}).\\

To deal with the stiffness problem arising due to the chemical reactions, a second-order accurate Strang operator splitting method~\cite{Strang1968} was employed to isolate this stiff source term. Equation~(\ref{N1}) was divided into two equations: Eq.~(\ref{N4}) a homogeneous partial differential equation for the fluid dynamics and Eq.~(\ref{N5}) an ordinary differential equation for the chemical reaction.
\begin{equation}
\frac{\partial \vec{U}}{\partial t}+\frac{\partial \vec{E}}{\partial x}+\frac{\partial \vec{F}}{\partial y}+\epsilon{\vec{S}}_{\mathrm{nr}}=0
\label{N4}
\end{equation}
\begin{equation}
\frac{\partial \vec{U}}{\partial t}={\vec{S}}_{\mathrm{r}}
\label{N5}
\end{equation}
Then, a finite-volume algorithm and a third-order TVD Runge-Kutta method~\cite{Shu1988} were used for spatial and temporal discretization for Eq.~(\ref{N4}). The AUSM+ scheme~\cite{Liou1993}, based on the fact of convection and acoustic waves as two physically distinct processes, was used to deal with the inviscid flux as a sum of the convective and pressure terms. The ordinary differential equation for the chemical reaction was solved using a fully implicit method.\\

The boundary condition along the $x$-axis was a mirror boundary condition (axis of symmetry), so that only the upper half of the layer is simulated in the case of a two-dimensional slab and axisymmetric cylinder. The upper boundary of the computational domain (above the inert layer) was a supersonic outflow condition to ensure that no reflected waves return into the computational domain. A study of the variation of the thickness of the inert layer confirmed that it was sufficiently thick to have no influence on the results reported below.\\

\section{Results}
\label{sec:results}
\subsection{Confinement of Equal Impedance of Explosive}
\label{sec:euqal}
\begin{figure}
	\centering
		\includegraphics[width=0.5\textwidth]{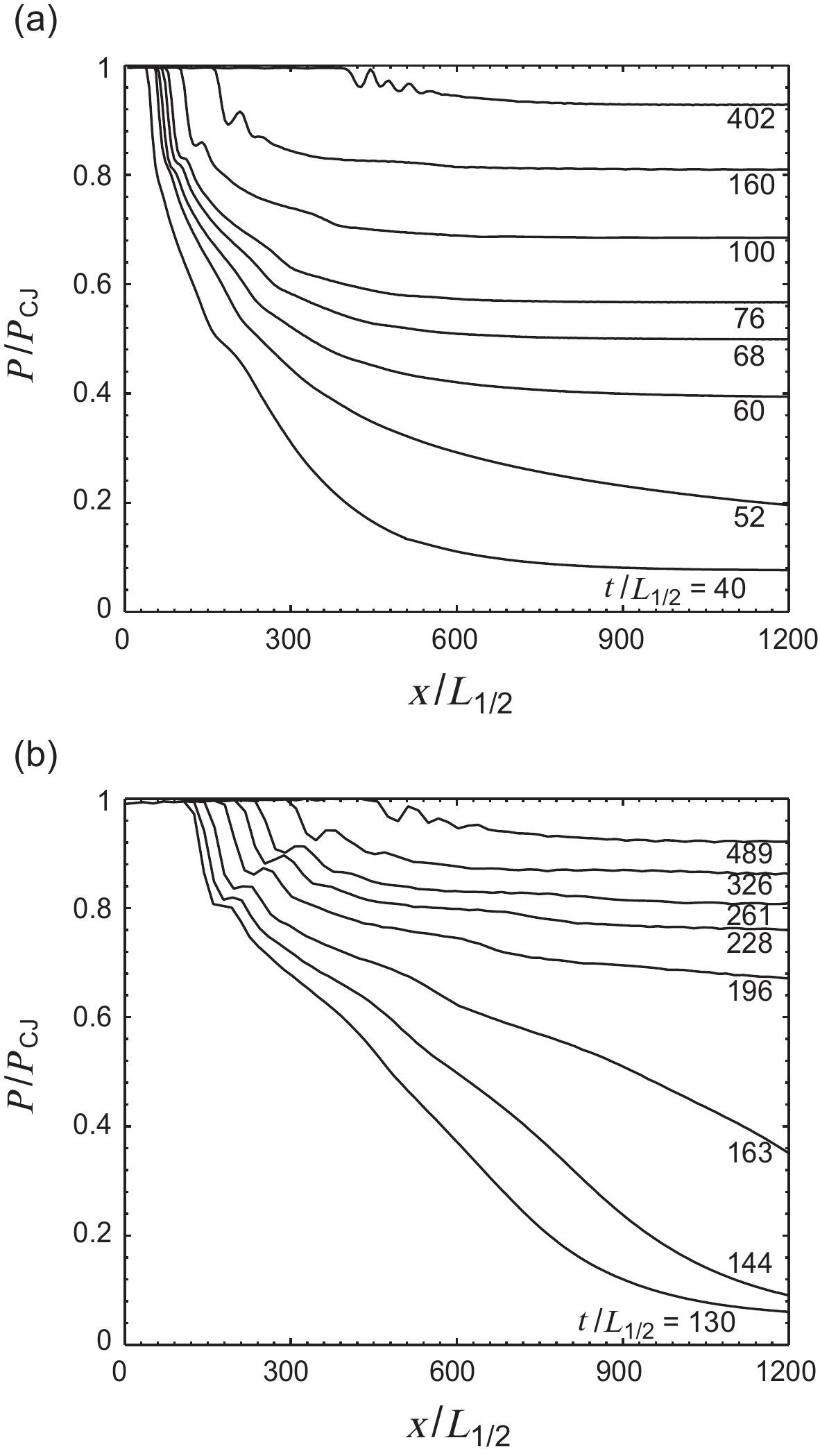}
		\caption{Detonation shock front pressure (normalized by the shock front pressure of an ideal CJ detonation) as a function of propagation distance (normalized by half reaction zone length), showing the decay of the detonation wave from its initial CJ velocity to the steady state terminal velocity, with (a) pressure exponent of $n=2$ and (b) pressure exponent of $n=3$.}
	\label{fig:Figure2}
\end{figure}

The results of simulations where the density of the inert layer was set equal to the density of the explosive (resulting in equal impedance) are shown in Fig.~\ref{fig:Figure2}, showing the pressure of the wave as a function of propagation distance for the two-dimensional, planar simulations. For each output data file, the location where the density along the central axis of the charge increased to twice the initial density of the explosive was considered as the position of the leading shock front at that time. Note the monotonic decay of the shock front from its initial ideal CJ value to its final, steady state velocity.  No oscillations of the shock front were observed, confirming that the pressure dependent reaction rate results in a stable detonation structure.  For the case of $n=2$, it was not possible to determine a critical thickness; steady propagation as always observed as the charge was made thinner. As the pressure exponent was made larger ($n=3$), a sharp cut-off between steady propagation or continued decay to failure was obtained, as discussed in greater detail in Section \ref{sec:pressure} below.  Similar behavior was obtained for the axisymmetric simulations. The rate of decay of the initially ideal CJ detonation to its steady state velocity with yielding confinement for all of the simulations performed in this study are further analyzed by Li et al.\cite{Li2014DS}\\

\begin{figure}[h]
	\centering
		\includegraphics[width=0.5\textwidth]{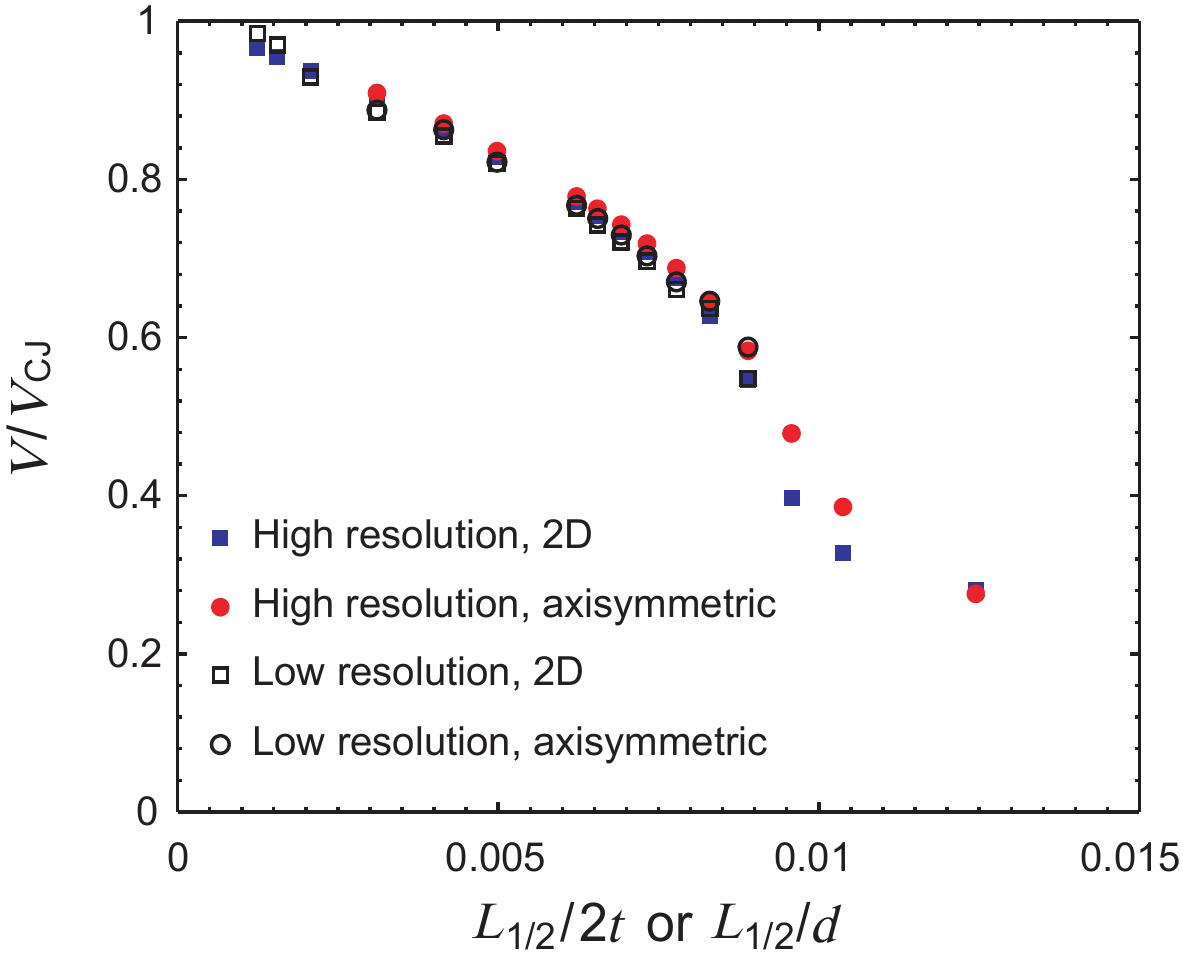}
		\caption{Detonation velocity (normalized by the ideal CJ velocity) for equal impedance confinement and $n=2$ as a function of the inverse charge diameter or thickness, normalized by the half reaction zone thickness. The thickness is scaled by a factor of two due to the expected $2:1$ scaling of diameter to thickness. Results for two different computational resolutions are shown, corresponding to 5 and 12.5 computational cells per half reaction zone thickness.}
	\label{fig:Figure3}
\end{figure}

The final steady state velocity is plotted in Fig.~\ref{fig:Figure3} as a function of the inverse thickness (for planar two-dimensional simulations) or inverse of the diameter (for axisymmetric simulations), with the detonation velocity normalized by the ideal CJ velocity.  The $x$-axis is plotted as the inverse of the diameter or thickness following the conventions of the condensed phase detonation literature, which plots data in this matter so that extrapolation to the $y$-axis yields the ideal CJ detonation velocity of an infinite diameter charge. The factor of two is included in the thickness in anticipation of the expected $2:1$ scaling between these two geometries. The dimension is also nondimensionalized by the half reaction zone thickness of the ideal CJ detonation.  A numerical convergence study was performed by increasing the computational grid resolution from 5 cells per half reaction zone length to 12.5 cells per half reaction zone length, which are reported as “low resolution” and “high resolution” results in Fig.~\ref{fig:Figure3}.  No difference was observed between these results, confirming that the simulations were performed with sufficient resolution to be effectively converged.  Note that this resolution is low in comparison to that typically used in simulations of detonations with Arrhenius kinetics, but due to the stable structure of the wave, this resolution was proven to be sufficient.  A resolution of 5 cells per half reaction zone is used for the rest of this study, although selected computations were repeated with 12.5 cells per half reaction zone length or greater in order to verify grid independence of the results.\\

\begin{figure}
	\centering
		\includegraphics[width=1.0\textwidth]{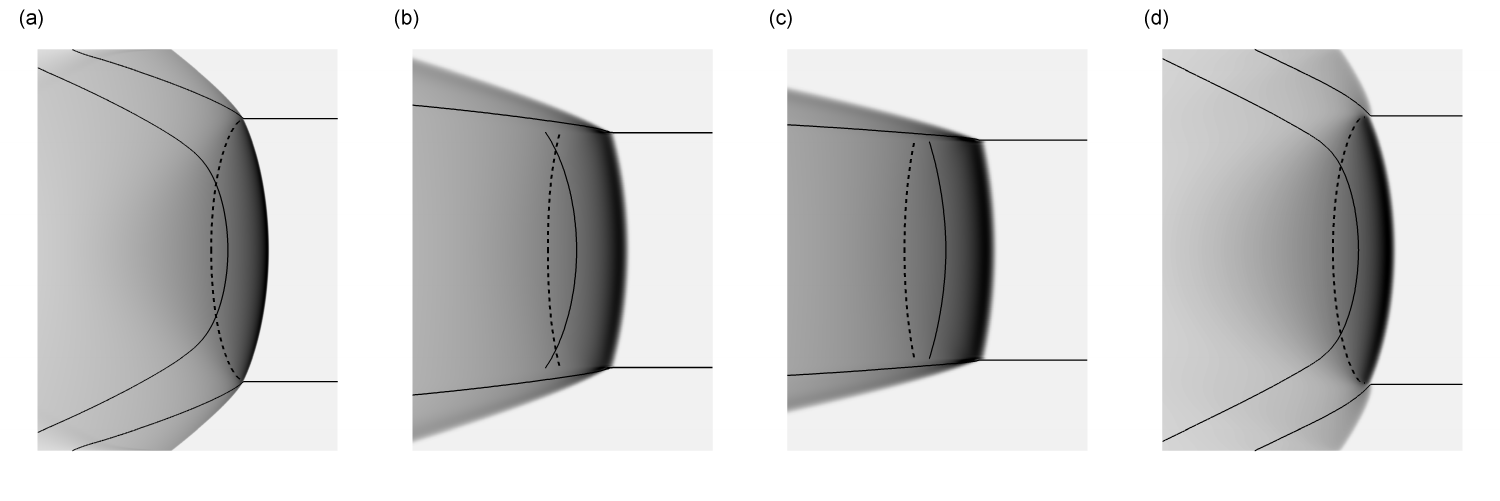}
		\caption{The structure of steady detonation waves obtained in two-dimensional (planar) numerical simulations with pressure exponent $n = 2$ for the cases of (a) confinement with density equal to explosive (b) confinement with density 5 times the explosive (c) confinement with density 10 times the explosive and (d) confinement with density half that of the explosive. The detonation propagates from left to right. The greyscale shading corresponds to pressure. The initially straight contour line is the boundary between the inert confinement and the reacting explosive. The dashed line is the sonic surface relative to the shock front. The contour behind the shock front is the location where the explosive has released 75\% of its chemical energy ($Z = 0.75$). All cases shown correspond to a slab thickness at which the detonation propagates at 75\% of the ideal detonation velocity.}
	\label{fig:Figure4}
\end{figure}

The structure of the steady wave obtained is shown in Fig.~\ref{fig:Figure4}~(a), showing the pressure field.  The sonic surface was found by transforming the results into a steady, wave-fixed frame using the terminal propagation velocity, and then normalizing the flow velocity relative to the shock front by the local speed of sound.  The detonation in the case of the confinement being equal in impedance to the explosive results in a lens-like structure of the curved shock front and oppositely curved sonic surface, which meet at the boundary between the explosive and confinement. A sonic region behind the leading shock and a continuous shock front (i.e., equal oblique shock angles in the inert and reactive regions) across the equal confinement interface can be predicted by a graphical construction of shock polars, as shown in Appendix \ref{sec:shockpolar}.\\

\subsection{High Impedance Confinement}
\label{sec:high}
In order to examine the effect of higher impedance confinement on the results, the density of the inert bounding gas was increased to $5 \rho_\mathrm{o}$ and $10 \rho_\mathrm{o}$ while maintaining the pressure constant, resulting in the acoustic impedance of the confinement increasing by factors of 2.24 and 3.15, respectively.\\

The structure of the steady propagating wave for these cases is shown in Fig.~\ref{fig:Figure4}~(b) and (c).  Note that the sonic surface is no longer attached to the shock front at the point where the shock impinges upon the confinement. At the high impedance confinement interface, a strong oblique shock with subsonic downstream flow in the reactive region and a weak oblique shock with supersonic downstream flow in the inert region are predicted by a simple shock polar analysis, as shown in Appendix \ref{sec:shockpolar}. The detachment of the sonic surface from the shock front for high impedance confinement predicted by polar analysis is confirmed by these computational results. The outward expansion of the products in the reaction zone is significantly reduced in comparison to the case of confinement with impedance equal to that of the explosive (Fig.~\ref{fig:Figure4}~(a)).\\

\begin{figure}
	\centering
		\includegraphics[width=1.0\textwidth]{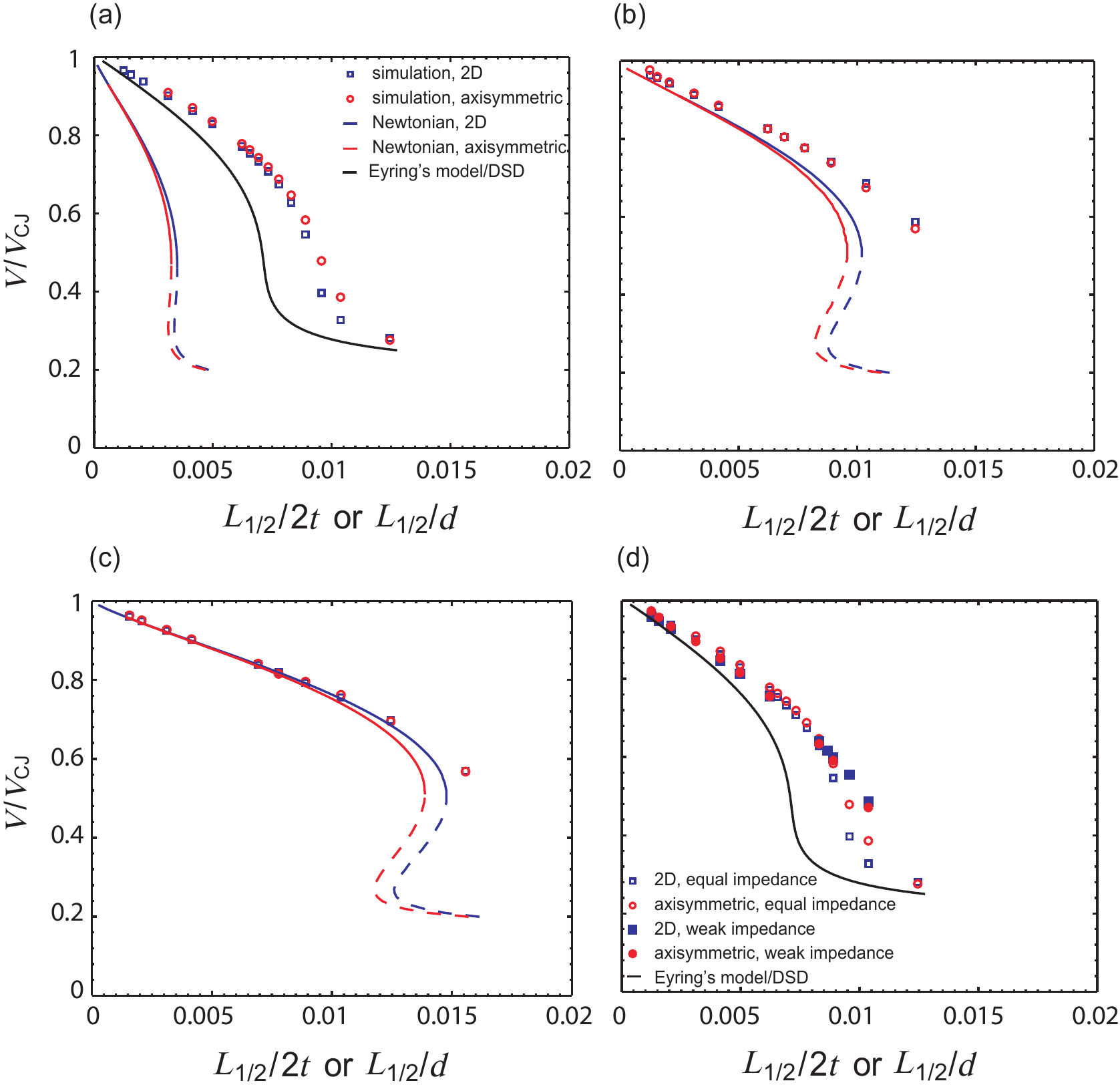}
		\caption{Detonation velocity (normalized by the ideal CJ velocity) as a function of the inverse charge diameter or thickness, normalized by the half reaction zone thickness, and compared to the front curvature model of Eyring et al.\cite{Eyring1949} and the Newtonian model of Tsuge et al.\cite{Tsuge1970} for a pressure exponent $n=2$ with the density of inert bounding gas equal to (a)~$\rho_\mathrm{o}$, (b)~$5 \rho_\mathrm{o}$, (c)~$10 \rho_\mathrm{o}$, and (d)~$0.5 \rho_\mathrm{o}$.}
	\label{fig:Figure5}
\end{figure}

The simulation results of the final steady state velocity are plotted as a function of the inverse thickness or inverse diameter for confinement density of $5 \rho_\mathrm{o}$ and $10 \rho_\mathrm{o}$ in Fig.~\ref{fig:Figure5}~(b) and (c), respectively. The $2:1$ scaling ratio between axisymmetric and two-dimensional geometries are verified by the simulation results for high impedance confinement. By quantitatively comparing the results for high impedance confinement to those for equal impedance confinement, the ability of higher impedance confinement to enable the detonation wave to propagate at greater velocity for the same charge thickness can be seen.
\subsection{Low Impedance Confinement}
\label{sec:low}
The effect of decreasing the impedance to a value less than the explosive was next examined. These simulations are salient in connection with propagation of bare solid explosive charges, which typically have a density more than three orders of magnitude greater than the ambient atmosphere.\\

As shown in Fig.~\ref{fig:Figure5}~(d), the simulation results of the final steady state velocity as a function of the inverse thickness or inverse diameter for low impedance confinement (i.e., confinement density of $0.5 \rho_\mathrm{o}$) are essentially indistinguishable to those for equal impedance confinement. In Fig.~\ref{fig:Figure4}~(d), the sonic surface has a lens-like shape attached to the shock/confinement impingement point, which is identical to that observed for the case of impedance confinement equal to that of the explosive in Fig.~\ref{fig:Figure4}~(a). The fact that there is no influence on the propagation or structure of the wave as the confinement density is decreased is consistent with the observation that the sonic surface is attached to the shock front at the confinement interface.  Once the sonic surface is attached to the shock/confinement impingement point, the structure of the wave is isolated from any further influence of the surrounding confinement.  In principle, the confinement density could be set to vacuum and the same wave speed and structure would be obtained.  This conclusion illustrates the relevance of these results to the impedance ratios that would be encountered with solid explosives with weak confinement or bare charges.
\subsection{Effect of Pressure Exponent}
\label{sec:pressure}
In order to examine the effect of this parameter, a series of calculations were performed with $n=3$. As seen in Fig.~\ref{fig:Figure2}~(b), this value resulted in a distinct separation between propagation and failure cases when examining the wave pressure as a function of distance, which was not observed in the case of $n=2$ (Fig.~\ref{fig:Figure2}~(a)). In Fig.~\ref{fig:Figure2}~(b) for $n=3$, the pressure-position curves corresponding to failure thicknesses decayed with an increasing magnitude of slope, while the successfully propagating cases asymptotically approached a final steady state.\\

\begin{figure}
	\centering
		\includegraphics[width=0.5\textwidth]{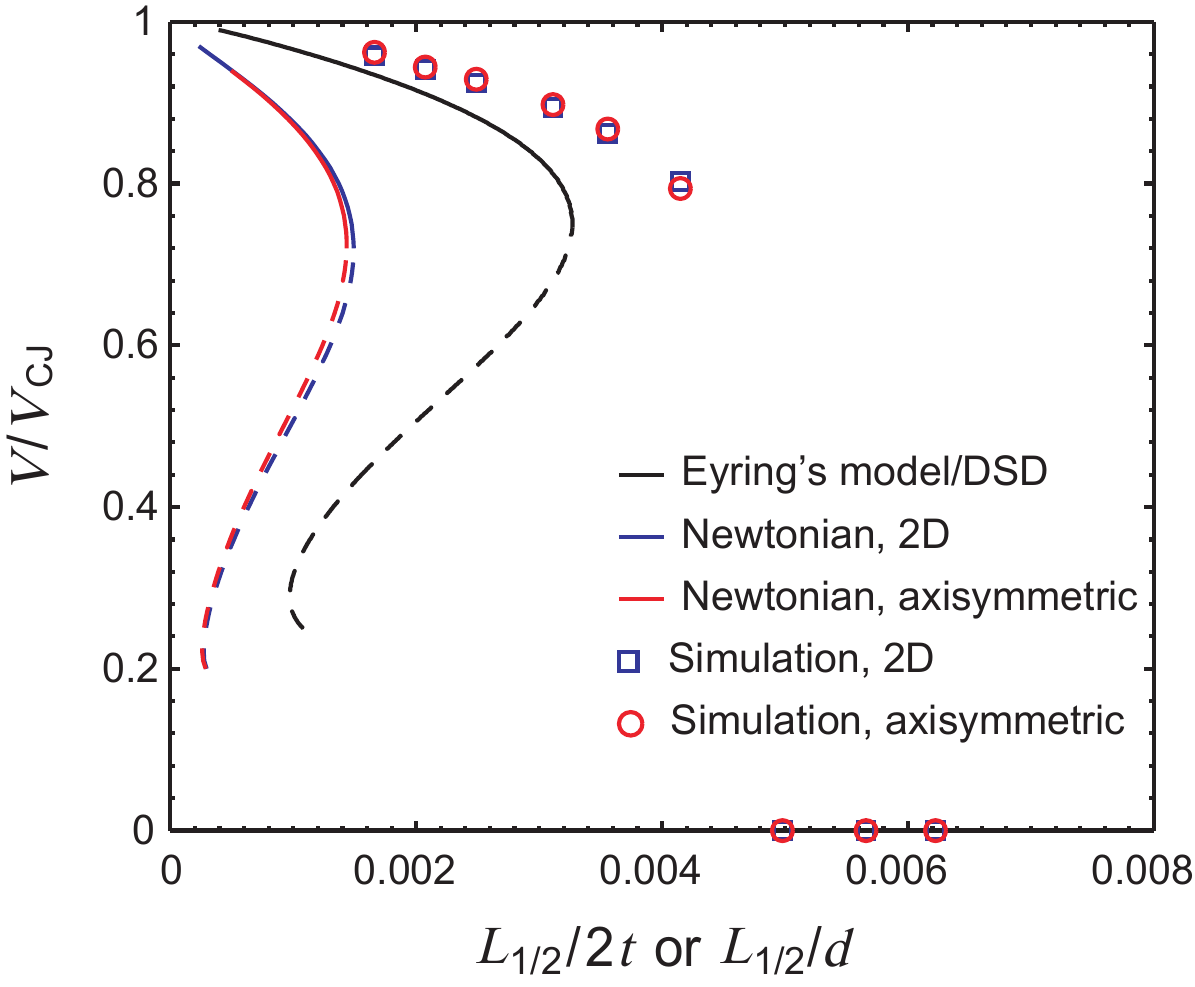}
		\caption{Detonation velocity (normalized by the ideal CJ velocity) as a function of the inverse charge diameter or thickness, normalized by the half reaction zone thickness, and compared to the front curvature model of Eyring et al.\cite{Eyring1949} and the Newtonian model of Tsuge et al.\cite{Tsuge1970} for pressure exponent $n=3$. Simulations that resulted in detonation wave failure are denoted with symbols along the $x$-axis.}
	\label{fig:Figure6}
\end{figure}

In Fig.~\ref{fig:Figure6}, the simulation results of steady state velocity for $n=3$ and equal impedance confinement are plotted as a function of the inverse thickness or inverse diameter. The scaling ratio $2:1$ is verified by comparing the results of two-dimensional and axisymmetric geometries for $n=3$.
\section{Analysis}
\label{sec:analysis}
\subsection{Normal Detonation Velocity and Shock Front Curvature Relation}
\label{sec:front}
The curvature of the leading shock front was measured for the steady state wave structures obtained in the simulations. A fourth-order polynomial function with respect to the transverse coordinate (i.e., $y$-direction) was fit to the extracted profile of the leading shock front. Since only the upper half of the charge was considered in the simulations, a complete profile of the leading shock front was obtained by plotting the mirror image of the upper-half profile below the centerline. Then, the curvature was evaluated at the centerline based on the fitting function.  The different geometries were taken into account by including the geometric factors corresponding to two-dimensional and axisymmetric geometries, 
\begin{equation}
	\kappa=\frac{1}{R_\mathrm{2D}}=\frac{2}{R_\mathrm{axisym}}
\end{equation}
where $R$ denotes the measured radius of curvature.\\

\begin{figure}
	\centering
		\includegraphics[width=0.5\textwidth]{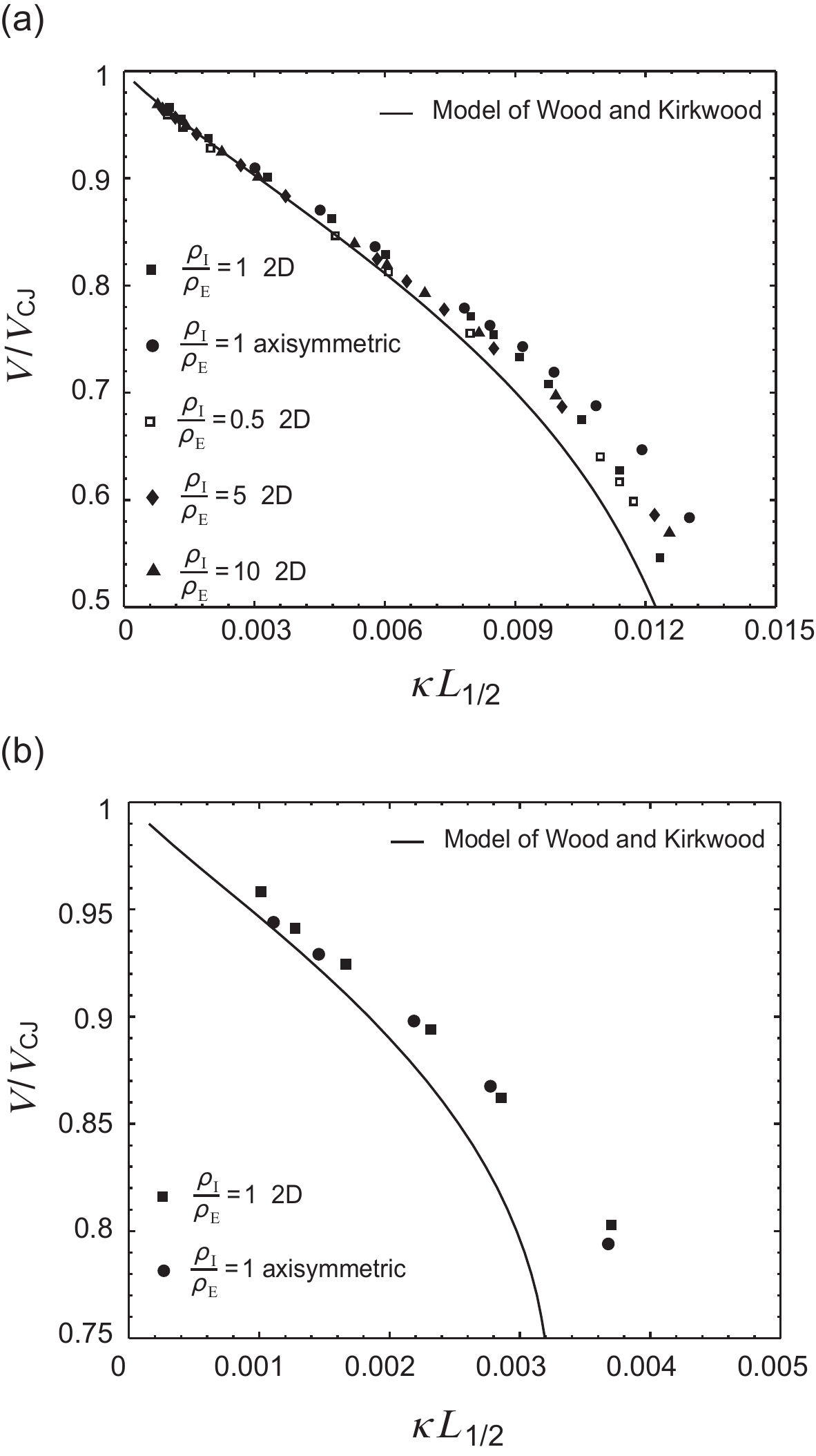}
		\caption{Front curvature measurement from the simulations and the ($D_\mathrm{N}$-$\kappa$) relation obtained by the model of Wood and Kirkwood\cite{WoodKirkwood1954} for (a) $n=2$ and (b) $n=3$}
	\label{fig:Figure7}
\end{figure}

In Figure~\ref{fig:Figure7}, the measured curvature of the leading shock front is plotted as a function of the inverse thickness or inverse diameter. Note that the results corresponding to low, equal, and high impedance confinements all fall onto a single curve. The simulation results exhibit good agreement with the normal detonation velocity and shock front curvature ($D_\mathrm{N}$-$\kappa$) relation obtained by solving the two-dimensional steady reactive Euler equations along the central axial streamline of the reaction zone.\cite{WoodKirkwood1954} More details on this calculation can be found in Appendix~\ref{sec:woodkirkwood}.
\subsection{Velocity Dependence on Charge Diameter or Thickness}
\label{sec:velocitythickness}
\subsubsection{Newtonian Model}
\label{sec:newtonianmodel}
In Figs.~\ref{fig:Figure5} and \ref{fig:Figure6}, the theoretical predictions of the Newtonian model of Tsuge et al.\cite{Tsuge1970} are plotted with the simulation results. This model treats the detonation with losses due to lateral expansion as a quasi-one-dimensional flow in a stream tube with the area divergence governed by a Newtonian theory-based treatment for the interaction with the confinement (complete details are provided in Appendix~\ref{sec:newtonian}). The curvature of the leading shock front is not explicitly considered in this model, i.e., a normal shock front is assumed. For confinements with equal or lesser impedance than the explosive, the radius of curvature is small in comparison to the charge dimension, meaning that the front is highly curved, relatively speaking, as can be seen in Fig.~\ref{fig:Figure4}~(a).  Unsurprisingly, as shown in Fig.~\ref{fig:Figure5}~(a) and Fig.~\ref{fig:Figure6}, the Newtonian model performs poorly in this case, over predicting the charge thickness or diameter at a given detonation velocity by a factor of approximately 4. In Fig.~\ref{fig:Figure5}~(b) and (c), the Newtonian model improves in agreement with the simulations as the impedance of the confinement is increased.  For the case with the density of the confinement equal to $10 \rho_\mathrm{o}$, the Newtonian model matches the computational results extremely well.  Examination of the wave structure in this case, as seen in Fig.~\ref{fig:Figure4}~(c), reveals a flow field that is amenable to a one-dimensional treatment.
\subsubsection{Front Curvature Model}
\label{sec:frontcurvature}
A more relevant model for the weak confinement cases incorporates a curvature-based geometric construction originally proposed by Eyring et al.\cite{Eyring1949} that utilizes the $D_\mathrm{N}$-$\kappa$ relation computed via a one-dimensional integration along the central streamline (Appendix~\ref{sec:woodkirkwood}).  In this model, any small portion of the wave front of the detonation propagating in a finite charge can be approximated by an infinitesimal segment of a steady spherical or cylindrical wave. The radius of this wave can be obtained from the $D_\mathrm{N}$-$\kappa$ relation providing the detonation velocity component normal to the local wave front, $D_\mathrm{N}$.\cite{Eyring1949} In original model of Eyring et al., the curved shock front was assumed to arrive parallel at the confinement boundary, but in our implementation, the critical oblique shock angle resulting in sonic flow downstream of the shock is used to terminate the construction and thus determine the critical diameter or thickness. The details of this geometric construction for leading shock front can be found in Appendix~\ref{sec:Eyring}. As shown in Fig~\ref{fig:Figure5}~(a), (d) and Fig.~\ref{fig:Figure6}, Eyring's construction makes a prediction that is in much better agreement with the numerical simulations than the Newtonian model for low and equal impedance confinements with both pressure exponent $n=2$ and $n=3$.\\

Modeling detonation wave fronts propagating with a curvature-dependent speed is the rationale for a formalism that is known as Detonation Shock Dynamics (DSD) theory.\cite{Bdzil1981, Bdzil1989} By adapting a front evolution equation methodology\cite{Stewart1989DetSymp, Aslam1996}, detonation velocity as a function of inverse charge thickness was calculated by relaxing a wave initialized as an ideal CJ detonation to its terminal, steady state configuration. The details of this calculation can be found in Appendix~\ref{sec:DSD}. It is worthwhile to point out that, given the appropriate $D_\mathrm{N}$-$\kappa$ relation and boundary shock angle, the theoretical predictions of the steady state wave shape made by DSD model are identical to that made by Eyring's construction. Hence, the results of the DSD model and Eyring's construction are plotted as one curve in Fig.~\ref{fig:Figure5}~(a), (d) and Fig.~\ref{fig:Figure6}.\\
\section{Conclusion}
\label{sec:conclusion}
The results of the present study confirm that the dimensional scaling of detonation velocity and critical dimension (diameter and thickness) between two-dimensional slab charges and axisymmetric cylindrical charges is very close to $2:1$ for the system considered in this paper, such that the critical diameter is encountered at twice the thickness of the critical slab.  In the case of weak confinement, the scaling is observed to be slightly less than $2:1$, meaning that the two-dimensional results lie slightly to the left of the axisymmetric results when plotted as $V/V_\mathrm{CJ}$ \textit{vs}. $1/d$ or $1/2t$. This result is qualitatively consistent with the findings of the analytical modelling of Bdzil\cite{Bdzil1981}, although the deviation from exact $2:1$ scaling obtained in the present study is never greater than $7\%$. This result is also consistent with the trends observed in recent measurements of condensed phase explosives made by Jackson and Short\cite{Jackson2013APS}, where the results with two-dimensional slab charges are found to lie to the left of results with axisymmetric cylindrical charges when plotted in this manner.\\

The issue of why certain highly heterogeneous explosives exhibit a scaling between diameter and thickness of greater than $2:1$ (i.e., the two-dimensional results lie to the right of the axisymmetric results when plotted as  $V/V_\mathrm{CJ}$ \textit{vs}. $1/d$ or $1/2t$)\cite{Petel2006DS, Petel2007JL, Higgins2009APS, Higgins2013APS, Jackson2013APS} remains unresolved by the present study.  The present results, which agree very well with classic front curvature models, lay a foundation for further investigation into the experimental findings of anomalous scaling.  An extension of the present work that introduced heterogeneity into the explosive media via a sinusoidal inhomogeneity has shown that large-scale heterogeneities can significantly influence propagation, allowing the detonation to propagate in significantly thinner slabs than the equivalent homogeneous case.\cite{Li2014CS}  The Li et al.\cite{Li2014CS} result appears to support the interpretation advanced by Petel et al.\cite{Petel2006DS} to explain their anomalous $4:1$ scaling results obtained with large-scale heterogeneity, namely, the mechanism of propagation becomes influenced by the local structure of the heterogeneous media.  The fact that a pressure-dependent reaction rate, resulting in a stable detonation wave structure, was used in the Li et al. study\cite{Li2014CS} allowed the effect of introducing heterogeneity to be clearly identified. Thus, the present study examining homogeneous explosive media comprises the “control” case of an ideal detonation that has been used in follow-on studies of non-ideal detonations.\\ 
   
The simple, curvature based construction proposed by Eyring et al.\cite{Eyring1949} does a remarkably good job in predicting the detonation velocity \textit{vs}. diameter/thickness relation, provided the shock interaction with the confinement is solved for (the original Eyring model assumed the shock was tangent to the confinement when it reached this boundary).  Provided the $D_{\mathrm{N}}$-$\mathrm{\kappa}$ relation obtained from Wood and Kirkwood's model is used, the prediction made by Eyring's construction exactly matches that made by DSD theory. The quasi-one-dimensional model of Tsuge et al.\cite{Tsuge1970}, which uses Newtonian theory to account for the interaction with the confinement, performs well in the case of high impedance confinement, wherein the sonic surface detaches from the shock/confinement interaction point and forms a downstream surface nearly parallel to the shock front.  In the case of weak confinement, however, where the detonation and sonic surface take on the lens-like structure, the Newtonian model performs poorly due to the inherent two-dimensional nature of the flow in the reaction zone following the curved shock front.  These models are highly valuable due to their transparency and simplicity in implementation.\\

\appendix

\section{Newtonian Model}
\label{sec:newtonian}
\begin{figure}
	\centering
		\includegraphics[width=0.5\textwidth]{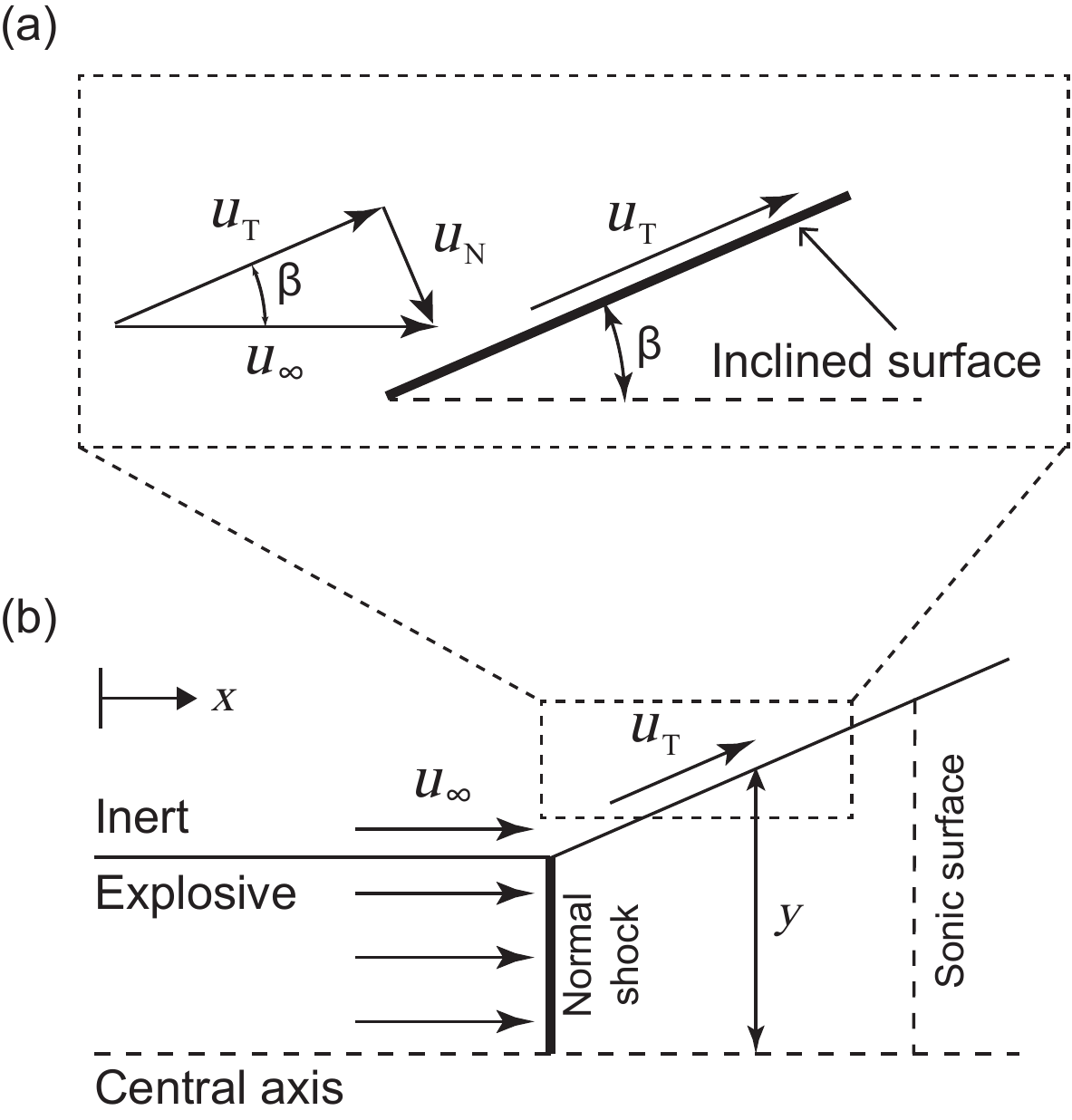}
		\caption{(a) Schematic for Newtonian model and (b) its implementation in the quasi-one-dimensional steady stream tube analysis}
	\label{fig:FigureA1}
\end{figure}

The Newtonian model assumes that, when a fluid flow encounters an inclined surface, it loses its momentum normal to the surface but conserves its tangential momentum. This model is accurate in predicting the pressure distribution on slender bodies in hypersonic flow.\cite{Anderson2000} As illustrated in Fig.~\ref{fig:FigureA1}~(a), the force $F$ exerted on the surface of area $A$ inclined at the angle $\beta$ by the impact of a stream with velocity $u_\infty$ is equal to the time rate of change of the momentum of this incoming flux,
\begin{equation}
\label{A1}
	\frac{F}{A}={\rho_{\infty}}{u_{\infty}}^{2}{\sin^{2} \beta}
\end{equation}
In the Newtonian model, the change in momentum flux of the flow, $\frac{F}{A}$, must be equal to the difference between the local static pressure on the surface $p$ and that in the freestream $p_{\infty}$, namely, 
\begin{equation}
\label{A2}
	\frac{F}{A}=p-p_\infty
\end{equation}
Hence, the local static pressure can be linked to the inclination angle $\beta$ of the surface via the following relation,
\begin{equation}
\label{A3}
	p=p_\infty+{\rho_{\infty}}{u_{\infty}}^{2}{\sin^{2} \beta}
\end{equation}
\\

In the steady quasi-one-dimensional stream tube model of detonation propagating in an explosive media confined by inert material, considering the hypersonic character of the confinement flow travelling towards the detonation wave, the Newtonian model can be used to approximate the flow divergence behind the leading shock. This approach was first proposed by Tsuge et al.\cite{Tsuge1970} The interaction between the flow of confinement material and the expansion of reacting explosive is treated by the Newtonian model as the incoming flow of confinement material striking an inclined surface as shown in Fig.~\ref{fig:FigureA1}~(b). The slope of the expanding stream tube surface $\frac{\mathrm{d}y}{\mathrm{d}x}$ is equivalent to the tangent of the local inclination angle $\beta$. In this steady quasi-one-dimensional stream tube analysis, the reacting and confinement materials must be in mechanical equilibrium, so the local static pressure in the reacting flow has to match that in the confinement flow. Hence, by replacing $\sin^{2} \beta$ in terms of $\frac{\mathrm{d}y}{\mathrm{d}x}$ in Eq.~(\ref{A3}), a relation linking the slope of the stream tube to the local static pressure $p(x)$ can be obtained as follows,
\begin{equation}
\label{A4}
	\frac{\mathrm{d}y}{\mathrm{d}x}=\sqrt{\frac{\frac{p(x)-p_\mathrm{I}}{\rho_{\mathrm{I}}{u_{\mathrm{I}}}^{2}}}{1-\frac{p(x)-p_\mathrm{I}}{\rho_{\mathrm{I}}{u_{\mathrm{I}}}^{2}}}}
\end{equation}
where subscript ``I'' denotes the properties of the confinement material before encountering the expanding reacting explosive and $y$ is either the radius or half-thickness of the charge in axisymmetric or two-dimensional slab geometries, respectively. The Newtonian model assumes a normal leading shock; the curvature of the detonation wave front is not considered explicitly by this model. Equation~\ref{A4} is then solved coupled to the conservation of mass, momentum, and energy for a quasi-one-dimensional stream tube initialized at the post-shock conditions. The complete formulation of the one-dimensional steady reactive Euler equations coupled with area divergence approximated by the Newtonian model can be found in Ref. 18.
\section{Shock Polar Analysis of Explosive and Inert Confinement Interface}
\label{sec:shockpolar}
Since the pressure and flow deflection immediately behind the leading shock must be matched at the interface separating the explosive and the inert confining material, a simple shock polar analysis can provide a description of the local interaction of the detonation products with the confinement. The application of shock polar analysis to classify different types of detonation confinements can be found in the literature.\cite{CourantFriedrichs} For a given upstream Mach number $M_\mathrm{x}$, the shock polar is the relation between the pressure ratio $\frac{p_\mathrm{y}}{p_\mathrm{x}}$  and the flow deflection angle $\delta$ across the shock. The oblique shock relations for ideal gas are used to determine $\frac{p_\mathrm{y}}{p_\mathrm{x}}$ and $\delta$ are,
\begin{equation}
\label{A5}
\frac{p_\mathrm{y}}{p_\mathrm{x}}=\frac{2\gamma{M_\mathrm{x}}^{2}{\sin^{2} \sigma}-(\gamma-1)}{\gamma+1}
\end{equation}
\begin{equation}
\label{A6}
\delta=\arctan {\frac{({M_\mathrm{x}}^{2}{\sin^{2} \sigma}-1)\cot \sigma}{\frac{\gamma+1}{2}{M_\mathrm{x}}^{2}-{M_\mathrm{x}}^{2}{\sin^{2} \sigma}+1}}
\end{equation}
where $\sigma$ is the oblique shock angle to the incoming flow and $\gamma$ is the heat capacity ratio. By fixing $M_\mathrm{x}$ and varying $\sigma$ from Mach angle to $90^\circ$, a shock polar can be constructed via Eqs.~(\ref{A5}) and (\ref{A6}) as a curve in $\frac{p_\mathrm{y}}{p_\mathrm{x}}$-$\delta$ space. The turning point on a shock polar is associated with the maximum flow deflection angle and, for an ideal gas, is very close to  the condition of sonic flow downstream. The branch of shock polar above the turning point corresponds to a strong oblique shock and subsonic downstream flow; the branch below the turning point corresponds to a weak oblique shock and supersonic downstream flow.\\

\begin{figure}
	\centering
		\includegraphics[width=0.5\textwidth]{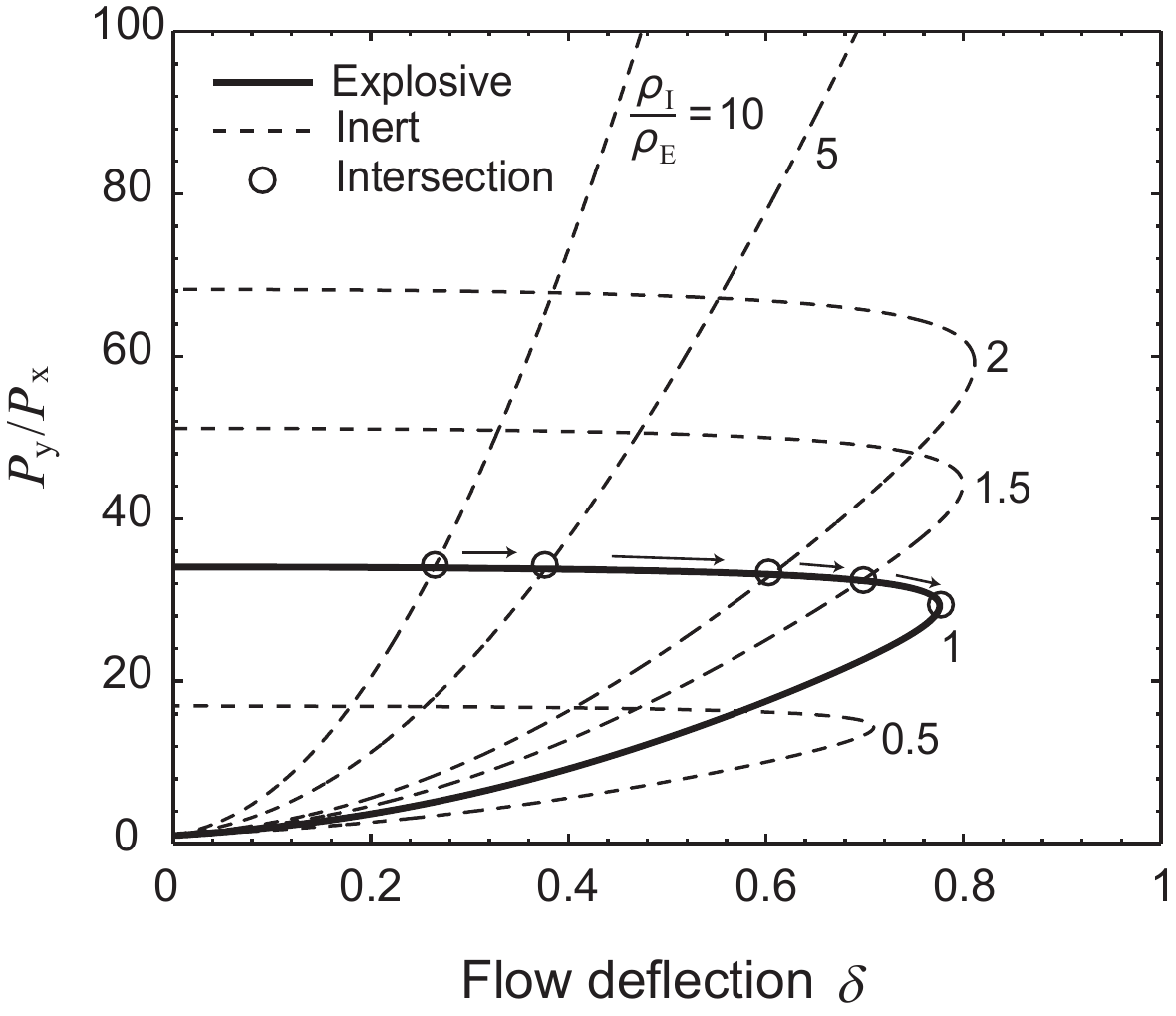}
		\caption{Shock polar curves of the explosive and inert confinement with various densities for incoming flow at the Chapman-Jouguet detonation velocity}
	\label{fig:FigureA2}
\end{figure}

In the current study, since the confinement material is modelled as the same gas as the explosive but with reaction progress variable equal to 1 (i.e., completely burnt), the acoustic impedance of the inert confinement is adjusted by inversely varying its initial density and temperature while maintaining the same initial pressure as that in the explosive. When the explosive/inert density ratio $\frac{\rho_\mathrm{I}}{\rho_\mathrm{E}}$  is unequal to unity, two different shock-polar curves are found in the explosive and inert regions. Hence, the shock polars corresponding to the confinement material with various initial densities and explosive material can be obtained as in Figure~\ref{fig:FigureA2}. The intersection of the explosive shock polar with each confinement shock polar gives the pressure and flow deflection matching conditions for each corresponding $\frac{\rho_\mathrm{I}}{\rho_\mathrm{E}}$.\\

As shown in Figure~\ref{fig:FigureA2}, when $\frac{\rho_\mathrm{I}}{\rho_\mathrm{E}}>1$, the intersection is on the strong branch of the explosive shock polar and weak branch of the confinement shock polar. This predicts that, for strong confinement, a subsonic region exists behind the leading shock in the explosive, namely, the sonic line is detached from the detonation wave front. When $\frac{\rho_\mathrm{I}}{\rho_\mathrm{E}}=1$, the shock polar curves of confinement and explosive coincide with each other, so no unique intersection can be identified directly. However, as shown in Figure~\ref{fig:FigureA2}, in the limit as $\frac{\rho_\mathrm{I}}{\rho_\mathrm{E}}$ decreases to unity, the intersection point approaches the turning point of the explosive shock polar, which implies that a sonic downstream condition must be satisfied when $\frac{\rho_\mathrm{I}}{\rho_\mathrm{E}}=1$.  This permits the shock angle in the explosive and confinement to be uniquely determined for the case of equal density confinement.

\section{Normal Detonation Velocity and Shock Front Curvature Relation}
\label{sec:woodkirkwood}
As Wood and Kirkwood\cite{WoodKirkwood1954} originally proposed, the $D_\mathrm{N}$-$\kappa$ relation can be obtained by solving the two-dimensional steady reactive Euler equation along the central axial streamline of the reaction zone. Given conditions of symmetry along this streamline, the transverse or radial flow velocity $v$ is zero. Hence, the continuity equation along this streamline can be written as,
\begin{equation}
\label{A7}
\frac{\partial \rho u}{\partial x}+\alpha\rho\frac{\partial v}{\partial y}=0
\end{equation}
where $y$ denotes the transverse and radial coordinate, and $\alpha$ has the value of 1 and 2 for the two-dimensional slab and axisymmetric geometries, respectively.\\

By performing a simple geometrical analysis along the leading shock front, a relation between the derivative of radial flow immediately behind the shock and the curvature of the shock front at the central axis can be obtained as,
\begin{equation}
\label{A8}
\frac{\partial v}{\partial y}=\frac{D-u_\mathrm{vN}}{R}
\end{equation}
where $u_\mathrm{vN}$ is the axial velocity at the von Neumann point, and $R$ is the local radius of curvature at the central axis. Considering that the length of the effective reaction zone is much smaller than the radius of curvature of the shock front,  $\frac{\partial v}{\partial y}$ can be assumed to only vary with local axial velocity while $R$ is taken as constant,
\begin{equation}
\label{A9} 
\frac{\partial v}{\partial y}=\frac{D-u(x)}{R}
\end{equation}
The shock radius of curvature $R$ can be linked to the shock front curvature $\kappa$ through the following relation,
\begin{equation}
\label{A10}
\kappa=\frac{\alpha}{R}
\end{equation}
By numerically solving the two-dimensional steady reactive Euler equations coupled with the relation between the shock front curvature and the derivative of radial flow velocity along the central axial streamline, a unique value of $\kappa$ can be determined for a given detonation velocity. Thus, detonation velocity can be obtained as a function of shock front curvature. This relation relates the normal component of detonation velocity $D_\mathrm{N}$ to the local shock front curvature away from the central axis.  Note that the $\alpha$ appearing in Eq.~(\ref{A10}) will cancel out with that in the equations of mass and momentum conservation, so the solutions of the governing equations for the $D_\mathrm{N}$-$\kappa$ relation for two-dimensional slab and axisymmetric geometries will be the same. This is verified in Fig.~\ref{fig:Figure7} in the main text of this paper.

\section{Geometric Construction of Detonation Wave Front}
\label{sec:Eyring}
\begin{figure}
	\centering
		\includegraphics[width=0.5\textwidth]{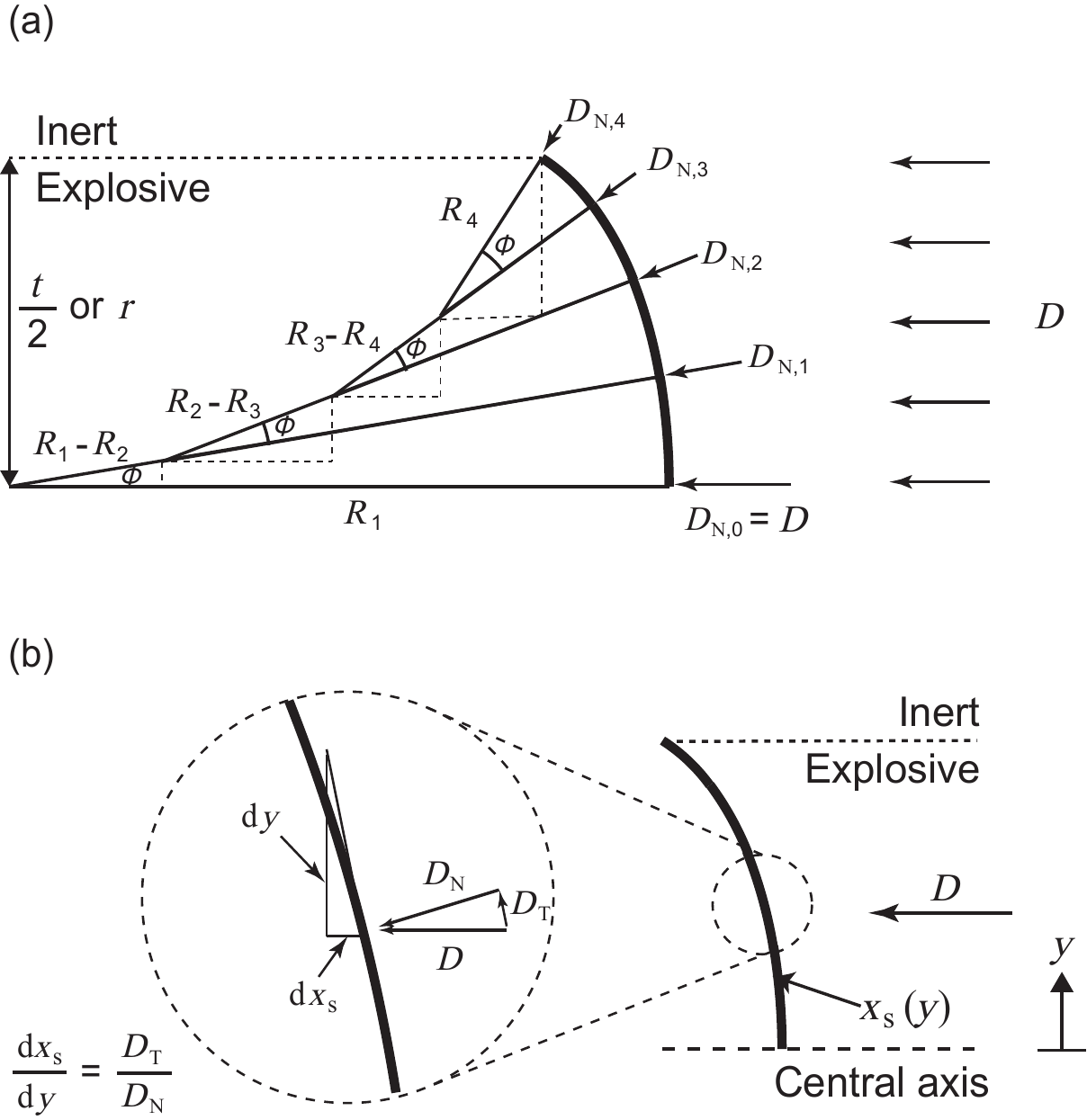}
		\caption{Illustations of (a) the geometric construction of detonation wave front and (b) geometric relation between local $D_\mathrm{N}$ and the slope of the shock wave front profile}
	\label{fig:FigureA3}
\end{figure}

The approach of geometrically constructing the detonation wave front utilizing the $D_\mathrm{N}$-$\kappa$ relation (see Appendix~\ref{sec:woodkirkwood}) was originally developed by Eyring et al.\cite{Eyring1949} to obtain a relation for the detonation velocity as a function of charge thickness or radius. In this model, any small portion of the wave front of the detonation propagating in a finite sized charge can be approximated by an infinitesimal segment of a steady spherical or cylindrical wave.\cite{Eyring1949} The radius of this wave can be obtained from the $D_\mathrm{N}$-$\kappa$ relation providing the detonation velocity component normal to the local wave front, $D_\mathrm{N}$. As illustrated in Figure~\ref{fig:FigureA3} (a), starting from the central axis, a series of arc segments, which have their sweep angles of a small incremental angle $\phi$ and radii determined through the obtained $D_\mathrm{N}$-$\kappa$ relation, are piecewise drawn and connected to define the shape of the detonation wave front. From this simple geometrical analysis, the following relation can be obtained to calculate the half-thickness or the radius of the charge,
\begin{equation}
\label{A11}
\frac{t}{2} \mathrm{~or~} r=R_{n}{\sin n\phi}+\sum\limits_{i=1}^{n-1} (R_{i}-R_{i+1})\sin {i\phi}
\end{equation}
where $n$ denotes the total number of incremental angle $\phi$'s required to reach the boundary of the charge. To find $n$, the shock polar matching condition needs to be applied at the explosive and confinement boundary. As the shock front angle to the incoming flow is obtained at the confinement boundary, $n$ can be calculated via the following relation,
\begin{equation}
\label{A12}
n\phi=90^\circ-\sigma_\mathrm{B}
\end{equation}
Provided $\sigma_\mathrm{B}$, the shock front angle at the confinement boundary (see Appendix~\ref{sec:shockpolar}), and $D_\mathrm{N}$-$\kappa$ relation (see Appendix~\ref{sec:woodkirkwood}) are known, the approach of constructing the detonation wave front can be then applied to solve for the steady detonation velocity as a function of charge thickness or radius.\\

In the limit where the incremental angles are taken as infinitesimally small, the Eyring construction can be cast in the form of a differential equation. which relates the mathematical formulation of local curvature in terms of local wave front position, $x_{\mathrm{s}}(y)$, and its first and second-order derivatives with respect to $y$, to the local curvature determined by the calculated $D_\mathrm{N}$-$\kappa$ relation. As illustrated in Fig.~\ref{fig:FigureA3}~(b), the local $D_\mathrm{N}$ can be related to the local slope of the wave front profile, $\frac{\mathrm{d} x_{\mathrm{s}}}{\mathrm{d} y}$, as follows,
\begin{equation}
\label{A13}
D_{\mathrm{N}}=\frac{D}{\sqrt{1+\left( \frac{\mathrm{d} x_{\mathrm{s}}}{\mathrm{d} y} \right)^2}}
\end{equation}
Thus, the governing differential equation can be formulated as,
\begin{equation}
\label{A14}
\frac{\frac{\mathrm{d}^2 x_{\mathrm{s}}}{\mathrm{d} y^2}}{\left[ 1+\left( \frac{\mathrm{d} x_{\mathrm{s}}}{\mathrm{d} y} \right)^2 \right]^{\frac{3}{2}}}=\kappa \left( D_{\mathrm{N}} \right)=\kappa \left( \frac{D}{\sqrt{1+\left( \frac{\mathrm{d} x_{\mathrm{s}}}{\mathrm{d} y} \right)^2}} \right)
\end{equation}
This equation can be numerically integrated starting from the centerline boundary condition ($\frac{\mathrm{d} x_{\mathrm{s}}}{\mathrm{d} y}=0$) toward the edge of the charge in order to compute the critical diameter for a given detonation velocity. This equation is seen to be identical to that obtained via Detonation Shock Dynamics when the front evolution equation is relaxed to a steady state solution, as seen in Eq.~\ref{E9} in Appendix~\ref{sec:DSD}.\\

\section{Propagation of Detonation Wave}
\label{sec:DSD}

Stewart and Bdzil extended the wave front evolution method to Detonation Shock Dynamics theory, and derived a partial differential equation that governs the propagation of detonation wave fronts in energetic materials.\cite{Stewart1989DetSymp} Aslam et al. provided a detailed numerical implementation of solving this governing PDE derived using level set methods.\cite{Aslam1996}\\

For a two-dimensional unsteady detonation, the front evolution method solves for a field function, $\mathrm{\Psi}(x,y,t)$, which depends on spatial coordinates and time. The detonation wave front is identified as the zero-level surface function of the field function, i.e., $\mathrm{\Psi}(x,y,t)=0$. Hence, the total derivative of this surface function is zero, i.e.,
\begin{equation}
\label{E1}
\frac{\mathrm{d} \mathrm{\Psi}}{\mathrm{d} t}=\frac{\partial \Psi}{\partial t}+\frac{\partial \Psi}{\partial x}\frac{\mathrm{d} x}{\mathrm{d} t}+\frac{\partial \Psi}{\partial y}\frac{\mathrm{d} y}{\mathrm{d} t}=0
\end{equation}
Equation~(\ref{E1}) can be expressed in terms of the surface velocity, $\vec{D}$, and gradient field of $\mathrm{\Psi}$, $\vec\nabla \mathrm{\Psi}$, as follows,
\begin{equation}
\label{E2}
\frac{\partial \Psi}{\partial t}+\vec\nabla \mathrm{\Psi} \cdot \vec{D}=0
\end{equation}
The unit vector $\hat{\mathrm{n}}$ is normal to the wave front surface and along the direction in which the wave propagates. The unit normal vector is given by $\hat{\mathrm{n}}=\vec\nabla \mathrm{\Psi}/| \vec\nabla \mathrm{\Psi} |$. Hence, the normal component of detonation velocity can be expressed as follows,
\begin{equation}
\label{E3}
\vec{D} \cdot \hat{\mathrm{n}}=\vec{D} \cdot \frac{\vec\nabla \mathrm{\Psi}}{| \vec\nabla \mathrm{\Psi} |}=D_\mathrm{N}(\mathrm{\kappa})
\end{equation}
Substituting Eq.~(\ref{E3}) into (\ref{E2}), we can obtain a Hamilton-Jacobi-like equation for the field function,
\begin{equation}
\label{E4}
\frac{\partial \Psi}{\partial t}+D_\mathrm{N}(\mathrm{\kappa})| \vec\nabla \mathrm{\Psi} |=0
\end{equation}\\

With applied boundary conditions and known $D_\mathrm{N}$-$\kappa$ relation, a relation between the local wave front curvature and the field function is required to obtain a closed-form PDE. This is achieved by explicitly expressing $\mathrm{\Psi}$ in the following form,
\begin{equation}
\label{E5}
\mathrm{\Psi}(x,y,t)=x-x_{\mathrm{s}}(y,t)-D_{\mathrm{CJ}}t
\end{equation}
where $D_{\mathrm{CJ}}$ is the Chapman-Jouguet detonation velocity. $\vec\nabla \mathrm{\Psi}$, $| \vec\nabla \mathrm{\Psi} |$, and $\frac{\partial \Psi}{\partial t}$ can be thus derived,
\begin{equation}
\label{E6}
\begin{split}
& \vec\nabla \mathrm{\Psi}=\frac{\partial \mathrm{\Psi}}{\partial x}\hat{\mathrm{e}}_{\mathrm{x}}+\frac{\partial \mathrm{\Psi}}{\partial y}\hat{\mathrm{e}}_{\mathrm{y}}=\hat{\mathrm{e}}_{\mathrm{x}}-\frac{\partial x_{\mathrm{s}}}{\partial y}\hat{\mathrm{e}}_{\mathrm{y}}\\
& | \vec\nabla \mathrm{\Psi} |=\sqrt{1+\left( \frac{\partial x_{\mathrm{s}}}{\partial y} \right)^2}\\
& \frac{\partial \Psi}{\partial t}=-D_{\mathrm{CJ}}-\frac{\partial x_{\mathrm{s}}}{\partial t}
\end{split}
\end{equation}
where $\hat{\mathrm{e}}_{\mathrm{x}}$ and $\hat{\mathrm{e}}_{\mathrm{y}}$ are the normal direction vectors of the laboratory-fixed reference frame. By substituting Eq.~(\ref{E6}) into (\ref{E4}), a PDE solving for $x_{\mathrm{s}}$ is obtained,
\begin{equation}
\label{E7}
\frac{\partial x_{\mathrm{s}}}{\partial t}=D_\mathrm{N}(\mathrm{\kappa})\sqrt{1+\left( \frac{\partial x_{\mathrm{s}}}{\partial y} \right)^2}-D_{\mathrm{CJ}}
\end{equation}
where, by definition, $\mathrm{\kappa}$ is
\begin{equation}
\label{E8}
\mathrm{\kappa}=\frac{\frac{\partial^2 x_{\mathrm{s}}}{\partial y^2}}{\left[ 1+\left( \frac{\partial x_{\mathrm{s}}}{\partial y} \right)^2 \right]^{\frac{3}{2}}}
\end{equation}\\

Equation~(\ref{E7}) can be equivalently solved in two ways to determine the relation between steady-state detonation velocity and charge thickness: (1) by fixing detonation velocity and applying its corresponding boundary conditionsand solving for the charge thickness; (2) by fixing charge thickness and guessing an initial detonation velocity, Eq.~(\ref{E7}) can be relaxed to a steady-state solution.\\

For the first method, by fixing the steady-state detonation velocity, $D$, Eq.~(\ref{E7}) can be reduced to an ordinary differential equation. Considering the geometric relation as illustrated in Figure~\ref{fig:FigureA3}~(b), this ODE can be formulated as
\begin{equation}
\label{E9}
D_{\mathrm{N}}(\mathrm{\kappa})=\frac{D}{\sqrt{1+\left( \frac{\partial x_{\mathrm{s}}}{\partial y} \right)^2}}
\end{equation}
The charge thickness that allows detonation to propagate at this given velocity can be determined by integrating Eq.~(\ref{E9}) from the central axial streamline to the location where the slope of the wave front matches the boundary oblique shock angle, $\mathrm{\sigma}_{\mathrm{B}}$. In fact, this way of solving the level set governing equation in DSD modeling is equivalent to Eyring's geometric construction taking the limit of the incremental angle, $\phi$, to zero, as can be seen by comparing Eqs.~(\ref{E8}) and (\ref{E9}) to (\ref{A14}).\\

In the second approach, Eq.~(\ref{E7}) has to be solved as a partial differential equation with fixed charge thickness. The detonation wave is initialized as a planar wave front propagating at the ideal CJ velocity.  $\mathrm{\sigma}_\mathrm{B}$ corresponding to this velocity can be calculated using the analysis presented in Appendix~\ref{sec:shockpolar}. By applying the calculated $\mathrm{\sigma}_\mathrm{B}$ as a time-invariant boundary condition, Eq.~(\ref{E7}) can be solved to obtain the relaxation process from the initially planar wave front to the steady-state curved wave front. Then, $\mathrm{\sigma}_\mathrm{B}$ can be recalculated using the new steady-state detonation velocity, and Eq~(\ref{E7}) can be re-solved applying the new $\mathrm{\sigma}_\mathrm{B}$. This calculation should be repeated till the difference between the steady-state detonation velocities obtained in two consecutive iterations is less than some numerical tolerance.\\

\begin{acknowledgments}
This work was supported by the National Natural Science Foundation of China (No. 51206150), National Key Laboratory for Shock Wave and Detonation Physics Research Foundation (No. 9140C6704010704), and the State Scholarship from China Scholarship Council.
\end{acknowledgments}

\bibliographystyle{unsrt}
\bibliography{detonation}

\end{document}